\documentclass{aastex631}

\usepackage{float}
\makeatletter
\let\newfloat\newfloat@ltx
\makeatother
\usepackage{amsmath,amssymb,amsfonts}%
\usepackage{amsthm}%
\usepackage{mathrsfs}%
\usepackage{algorithm}%
\usepackage{algpseudocode}%

\newtheorem{theorem}{Theorem}%  meant for continuous numbers

\newtheorem{corollary}{Corollary}[theorem]

\newtheorem{principle}{Principle}

\newtheorem{remark}{Remark}%

\newtheorem{definition}{Definition}%

 \newcommand*{\defeq}{\mathrel{\vcenter{\baselineskip0.5ex \lineskiplimit0pt
                     \hbox{\scriptsize.}\hbox{\scriptsize.}}}%
                     =}

\def\E{{\mathbf E}}
\def\I{{\mathbf I}}
\def\P{{\mathbf P}}
\def\V{{\mathbf V}}
\def\D{\mathrm{d}}

\begin{document}

\title{Is it possible to know cosmological fine-tuning?}

\author[0000-0001-6281-1720]{Daniel Andrés Díaz-Pachón}
\affiliation{University of Miami \\
1120 NW 14 ST Suite 1057 \\
Miami, FL 33136, USA}

\author{Ola H\"ossjer}
\affiliation{Stockholm University \\
SE-106 910 \\
Stockholm Sweden}

%\collaboration{20}{(AAS Journals Data Editors)}

\author{Calvin Mathew}
\affiliation{American Heritage School \\
12200 W Broward Blvd \\
Broward, FL 33325, USA}



\begin{abstract}
Fine-tuning studies whether some physical parameters, or relevant ratios between them, are located within so-called life-permitting intervals of small probability outside of which carbon-based life would not be possible. Recent developments have found estimates of these probabilities that circumvent previous concerns of measurability and selection bias. However, the question remains if fine-tuning can indeed be known. Using a mathematization of the concepts of learning and knowledge acquisition, we argue that most examples that have been touted as fine-tuned cannot be formally assessed as such. Nevertheless, fine-tuning can be known when the physical parameter is seen as a random variable and it is supported in the nonnegative real line, provided the size of the life-permitting interval is small in relation to the observed value of the parameter.
\end{abstract}



\section{Introduction}

Cosmological fine-tuning (FT) says that some physical parameters, some ratios between them, and some boundary conditions, must pertain to intervals of small probability to permit the existence of carbon-based life \citep{LewisBarnes2016}. FT started to spark the interest of the scientific community when \citet{Carter1974} brought it to light calling it the \textit{anthropic principle}, a term that, according to \citet{LewisBarnes2016}, was popularized and modified by \citet{BarrowTipler1988}. Since then, considerable effort has been made to determine the length of these so-called life-permitting intervals (LPIs), constituting the area where the scientific literature has primarily focused \cite[see, e.g.,][]{Adams2019, Barnes2012, Davies1982}. For this reason, some have linked FT with the small {\it length} of the intervals, with no regard for their probability \citep[see, e.g.,][]{Helbig2023}. This evokes the idea of ``naturalness'' in physics, where a good theory is assumed to present numbers that are close to unity, and departures from it point to ad-hoc explanations trying to fit observations to theory as Mercury epicycles in the Ptolemaic model \citep{Hossenfelder2020}. FT goes beyond naturalness, asserting that, according to our current theories, such unnaturalness is needed if carbon-based life is going to exist \citep{Barnes2021}. The existence of life sets a specification; i.e., a subset of possible outcomes of parameters for which life is permitted. These outcomes maximize a function $f$ that quantifies how specified outcomes are, and it is stochastically independent of the original random variable. In cosmological FT, this random variable is the physical parameter, whose value is assumed to be independent of the requisites for carbon-based life existence, and the LPI maximizes an indicator function $f$ that determines if life is possible or not. Therefore, the tuning of a physical constant for life is fine if and only if {\it the probability} of its LPI is small \citep{DiazHossjer2022}. 

Fine-tuning divides the scientific and philosophical communities between those who dismiss it as mere speculation \cite[e.g.,][]{Hossenfelder2020, ColyvanGarfieldPriest2005, McGrewMcGrewVestrup2001} and those who consider it a serious scientific endeavor \cite[e.g.,][]{Davies2008, LewisBarnes2016, Tegmark2015}. One of the main criticisms, even among those who see it as a legitimate question, is the lack of a valid probability distribution to impose over the parameters \citep{Adams2019}. In this direction, Hossenfelder asks: ``[H]ow do you make a statement about probability without a probability distribution?'' \cite[p.~205]{Hossenfelder2020}. The problem is indeed daunting: it requires estimating the probability of the LPI using a biased sample of size 1 taken from an unknown distribution supported in an unknown space. More explicitly, and to establish some notation, let $\mathcal X$ be the possible values of a random physical parameter $X$. FT is about inferring the probability of its LPI $\ell_X$ when the only observation we possess is $x_0$, the {\it observed} value of $X$ in our \textit{uni}verse; moreover, such an observation suffers from selection bias because that universe harbors life; and the only information about the probability distribution $F=F_X$ of $X$ is that $F$ is supported on $\mathcal X$ and $x_0\in\mathcal X$. Moreover, assuming Carter's definition, it is just natural to assume that $F$ corresponds to a prior distribution of $X$ that maximizes the entropy \citep{Carter1974}.

Past efforts to find the probabilities of LPIs use mainly uniform distributions over the set of possible values that constants might take \cite[see, e.g.,][]{Sandora2019a, Barnes2020, TegmarkEtAl2006}. Nonetheless, this approach attests to the difficulty of the task, since the choice of the uniform distribution is based on Bernoulli's Principle of Insufficient Reason \citep{Bernoulli1713}, which requires knowledge that the sample space is finite ---a strong assumption that cannot be warranted for cosmological FT. This legitimate criticism of FT measurements came from philosophy, where it was called the \textit{normalization problem} \citep{ColyvanGarfieldPriest2005, McGrewMcGrewVestrup2001, McGrewMcGrew2005, McGrew2018}. 

Thus, the tuning problem can be divided into two steps. First, determining the size of the LPI for a given constant of nature ---a task of physics---, and determining the probability of the LPI ---a mathematical task---. For this reason, we formally define FT next.

\begin{definition}[{\bf Fine-tuning}]\label{D:FT}
Fine-tuning happens if and only if $F_0(\ell_X)$ is small, where $F_0=F_X$ is the assumed true distribution of $X$, which is supported on $\mathcal X$. In more detail, there is fine-tuning if and only if there exists a real-valued $\delta >0$ such that
\begin{align}\label{FT}
    F_0(\ell_X)\le \delta \ll 1.
\end{align}
If the tuning is not fine, we will call it \textit{coarse}.
\end{definition}

Recently Díaz-Pachón, H\"ossjer, and Marks developed Algorithm \ref{JesusLovesU} (below) to find an upper bound on the probability of the {\it known} LPI $\ell_X$ of a parameter $X$ \citep{DiazHossjerMarks2021, DiazHossjerMarks2023}. Algorithm \ref{JesusLovesU} can be viewed as a statistical test, where the null hypothesis $H_0$ that tuning of $X$ is coarse is tested against the alternative hypothesis $H_1$ that $X$ is fine-tuned. The first two steps of Algorithm \ref{JesusLovesU} ask for the input: a parameter space $\mathcal X$ (e.g., $\mathbb N$, $\mathbb R^+$, $\mathbb R$, $\mathbb R^n$, $\mathbb C$, or some subset thereof), and the constraints to impose on the prior distributions $F_X$ supported on $\mathcal X$ (e.g., if some event $B \subset \mathcal X$ has a known probability, if some of the moments of $X$ are finite, etc.). In the third step, a collection 
\begin{align*}
    \mathcal F \defeq \{F(\cdot ; \theta); \theta \in\Theta \} 
\end{align*}
of all the possible maximum entropy (maxent) priors $F$ is determined that satisfy these constraints. These priors are indexed by a hyperparameter $\theta$ that varies over a hyperparameter space $\Theta$. In particular, we will assume that the true distribution of $X$ satisfies  
\begin{align}\label{obsF}
    F_0 = F(\cdot;\theta_0)\in \mathcal F
\end{align}
for some $\theta_0\in\Theta$. In the fourth step of Algorithm \ref{JesusLovesU} the maximum probability $\text{TP}_{\max}$ of the LPI $\ell_X$ is found over the collection of all the posterior probabilities $\{F(\ell_X; \theta); \theta\in\Theta\}$ induced by the priors in $\mathcal F$. That is, for each $\theta \in \boldsymbol\Theta$ we first define the corresponding marginal probability of $\ell_X$ as
\begin{align*}
    F(\ell_X; \theta) = \int_{\mathcal X} P(\ell_X \mid x) F(\D x; \theta),
\end{align*}
by integrating over all possible outcomes $x$ of $X\sim F(\cdot; \theta)$. We refer to $F(\ell_X; \theta)$ as a tuning probability. By maximizing this tuning probability over $\Theta$, the maximum tuning probability 
\begin{align*}
    \mathrm{TP}_{\max} \defeq \max_{\theta \in \Theta} F(\ell_X; \theta)
\end{align*}
is obtained. In the fifth step, the outcome of Algorithm 1 is produced: if $\text{TP}_{\max}$ is small ($\le\delta$ according to Definition \ref{D:FT}), we reject the null hypothesis $H_0$ and conclude there is FT. If $\text{TP}_{\max}$ is not small, the algorithm is inconclusive. In this case, we cannot reject $H_0$. This does not mean that $X$ is not finely tuned though. Indeed, $H_1$ may still be true when $H_0$ is not rejected, since $F_0(\ell_X)$ may be larger or smaller than $\delta$ when $\mathrm{TP}_{\max}>\delta$. 

\begin{algorithm}
\caption{Algorithm for testing fine-tuning}\label{JesusLovesU}
\begin{algorithmic}
	\item \textbf{Input:} Choose the set of possible values $\mathcal X$ of $X$. 
        \item \textbf{Input:} Define constraints on the distribution $F=F_X$ of $X$.
	\item Find the family ${\cal F}=\{F(\cdot;\theta);\theta\in\Theta\}$ of maxent distributions $F$ with support $\mathcal X$, subject to the constraints in Step 2, and assume ($H_0$) that the true distribution of $X$ satisfies $F_0=F(\cdot;\theta_0)\in \mathcal F$ for some $\theta_0\in\Theta$.
	\item Find the maximum tuning probability $\mathrm{TP}_{\max} \defeq \max \{F(\ell_X); \, F\in {\cal F}\}$ over $\mathcal F$ for the life-permitting interval $\ell_X$.
	\item \textbf{Output:} If $\mathrm{TP}_{\max}$ is small ($\le\delta$) there is fine-tuning (reject $H_0$); otherwise, the test is inconclusive (do not reject $H_0$).
\end{algorithmic}
\end{algorithm}

Algorithm \ref{JesusLovesU} has some important properties. First, the upper bound $\mathrm{TP}_{\max}$ is sharp because if the family ${\cal F}$ is made of a single maxent distribution $F_0$, then  $\text{TP}_{\max}= F_0(\ell_X)$. For instance, assume that Step 1 selects $\mathcal X=\mathbb R^+$, and in Step 2 it is assumed that the first moment is finite {\it and} known (e.g., $\E_0 X=\theta_0$). Then the family of Step 3 is made of a single distribution with density $F^\prime(x;\theta_0) = f(x;\theta_0) = e^{-x/\theta_0}/\theta_0$, the exponential with mean $\theta_0$, since this distribution is maxent among all the distributions with mean $\theta_0$ that are supported in $\mathbb R^+$. Consequently, $\text{TP}_\text{max}=F_0(\ell_X) = \int_{\ell_X} e^{- x/\theta_0}\text dx/\theta_0$.

Second, the selection of the class $\mathcal F$ as a family of maxent distributions simultaneously solves both the normalization and selection bias problems \cite[for a detailed discussion, see][]{DiazHossjerMarks2023}, maybe the two biggest concerns raised by physicists \citep{TegmarkEtAl2006, Adams2019, Hossenfelder2019, Hossenfelder2020} and philosophers \citep{ColyvanGarfieldPriest2005, McGrewMcGrewVestrup2001, Bostrom2002} alike. On one side, considering more general classes of maxent distributions than uniform ones solves the normalization problem. At the same time, by considering the whole class $\mathcal F$, and not only the distribution induced by an estimate of $\theta_0$, from the observed value of $X$ in our universe, the selection bias problem is solved. Consider for instance the case when $\mathcal X = \mathbb R^+$ and $\mathcal F$ is the class of exponential distributions with expected value $\theta\in \mathbb R^+$. It can then be shown that the maximum likelihood estimator of $\theta_0$ is $\hat{\theta}_0=X$. But we still have no guarantee that $F_0(\ell_X) = F(\ell_X; \theta_0)\le F(\ell_X; \hat{\theta}_0)$. On the other hand, by \eqref{obsF} we know that $F_0(\ell_X)\le \text{TP}_\text{max}$. Thus, considering all possible values of $\theta$ removes the bias induced by the weak anthropic principle, since the class $\mathcal F$ includes $F(\cdot, \theta_0)$ but is not reduced to it. 

Third, Algorithm \ref{JesusLovesU} reveals that the output heavily depends on the input, namely the sample space $\mathcal X$ for $X$ and the constraints to impose on the family ${\cal F}$ of prior distributions $F$ of $X$. Moreover, Díaz-Pachón, H\"ossjer, and Marks proved a  theorem  (Appendix 3 of \citet{DiazHossjerMarks2023}, summarized in Table \ref{T:Tuning} below, and explained in Remark \ref{Explan}) where they show that subtle changes in the input might produce extremely different outcomes. This can be seen, for instance, from Rows 4-5 of Table \ref{T:Tuning}, when considering families $\cal F$ of distributions over $\mathcal X=\mathbb R$ that depend on a scale parameter only: if $0\in \ell_X$, it will provoke $\text{TP}_{\text{max}}=1$, regardless the size of the interval $\ell_X$ (therefore the level of tuning cannot be assessed), whereas $0\notin\ell_X$ with midpoint $x_0$ of $\ell_X$, will produce a small $\text{TP}_{\text{max}}$, provided the relative half size $\epsilon=|\ell_X|/(2x_0)$ of the interval is small. In the same direction, when $\mathcal X = \mathbb R^+$ ($\mathcal = \mathbb R$) and the form and scale (location and scale) family $\mathcal F$ is considered, $\text{TP}_{\text{max}}$ jumps from very small to 1, depending on whether the maximal signal-to-noise ratio $T$ of the prior distributions $F\in\mathcal F$ in Rows 2-3 (7-8) is bounded or not. 

\begin{remark}\label{Explan}
    The results of Table \ref{T:Tuning} are presented for LPIs of the form
    \begin{align}\label{D0LPI}
        \ell_X = x_0[1-\epsilon,1+\epsilon],
    \end{align} 
    where $\epsilon$ is a dimensionless quantity representing the relative half size of the interval, and $x_0$ is the observed value of $X$, taken to be the middle point of the interval (if $x_0$ does not coincide with the middle point of $\ell_X$, nothing is lost by assuming the two values coincide and the notation is greatly simplified). When $\epsilon>0$ is small, fine-tuning probabilities are computed as a function of $\epsilon$ for diverse parametric families $\cal F$ of prior distributions $F$ of a randomly generated universe $X$. These $\mathrm{TP}_{\max}$ are presented given certain constraints on $\ell_X$ and/or $\theta\in\Theta$, the latter of which includes one or two variable hyperparameters of the prior densities $f(x;\theta)=F^\prime(x;\theta)$ in $\cal F$. The different choices of $\cal F$ correspond to a prior density $f(x/\theta)/\theta$ for the scale family with scale parameter $\theta$, a prior density $f(x/\theta_2;\theta_1)/\theta_2$ for the form and scale parameter with form parameter $\theta_1$ and scale parameter $\theta_2$, and a prior density $f((x-\theta_1)/\theta_2)/\theta_2$ for the location and scale family with location parameter $\theta_1$ and scale parameter $\theta_2$. When $\mathcal F$ involves two hyperparameters, the constraints on these hyperparameters are formulated in terms of the maximal value $T$ of the signal-to-noise ratio $\text{SNR}=E^2(X)/\mbox{Var}(X)$, that is, the maximal value of the ratio of the squared first moment and variance of the distribution $F$. There is also a constant associated to each family ${\cal F}$; $C_1=\max_{x >0} x f(x)$ for the scale family, $C_2=1/\sqrt{2\pi}$ for the form and scale family, and $C_3=\max_{x\in\mathbb{R}} f(x)$ for the location and location-scale families.
\end{remark}

\begin{table}[t]
\centering
\caption{Maximal Tuning Probabilities $\text{TP}_{\text{max}}$ 
\label{T:Tuning}}
\begin{tabular}{ccccc}
\hline
$\mathcal X$ & ${\cal F}$ & $\Theta$ & Constraint &  $\text{TP}_{\text{max}}$  \\
\hline
& Scale & $\mathbb R^+$ & $\epsilon \ll 1$ & $2\epsilon C_1$ \\
$\mathbb R^+$ & Form and scale & $\mathbb R^+\times \mathbb R^+$ & None & 1\\
& {\bf Form and scale} & $\mathbb R^+\times \mathbb R^+$ & $\text{SNR} \le T$, $T \gg 1$, $\epsilon \ll 1/\sqrt{T}$ &  $2\epsilon C_2\sqrt{T}$ \\
\hline
 & Scale & $\mathbb R^+$ & $0\notin \ell_X$, $\epsilon \ll 1$ & $2\epsilon C_1$\\
& Scale & $\mathbb R^+$ & $0\in \ell_X$ & 1 \\
$\mathbb R$ & Location & $\mathbb R$ & $\epsilon \ll \min(1/C_3,1)$ & $2\epsilon C_3$\\
& {\bf Location and scale}  & $\mathbb R\times \mathbb R^+$ & $\text{SNR} \le T$, $\epsilon \ll \min(1/\sqrt{T},1)$ & $2\epsilon (C_3\sqrt{T} + C_1)$\\
& Location and scale  & $\mathbb R\times \mathbb R^+$ & None & 1\\
\hline
\end{tabular}%
\end{table}

\section{Mathematical Framework for Learning and Knowledge Acquisition}\label{S:LKA}

The No Free Lunch theorems \citep{WolpertMacReady1995,WolpertMacReady1997} assert that, on average, no search does better than a blind one, and therefore a guided search infuses information to the search problem. Active information was thus defined to measure the amount of information introduced by a programmer in order to reach a target, compared to a baseline distribution which is usually, but not necessarily, in maxent \citep{DembskiMarks2009b, DiazMarks2020a}. Formally, active information is defined as
\begin{align}\label{AIN}
    I^+(A)\defeq \log \frac{\P(A)}{\P_0(A)},
\end{align}
where the target $A\subset\Omega$ is a subset of a search space $\Omega$, whereas $\P$ and $\P_0$ are probability distributions on $\Omega$ that represent searches of the programmer and blind search, respectively.  

Since its inception, active information has been used in the measurement of bias for machine learning algorithms \citep{Montanez2017a, Montanez2017b, MontanezEtAl2019, MontanezEtAl2021}, hypothesis testing \citep{DiazSaenzRao2020, HomYikMontanez2023}, statistical genetics \citep{ThorvaldsenHossjer2023, DiazMarks2020b}, bump hunting \citep{DiazEtAl2019, LiuEtAl2023}, and estimation and correction of prevalence estimators of Covid-19 \citep{HossjerEtAl2023, ZhouEtAl2023}, among others. In fact, active information can be used as a measure of FT if the search space $\Omega$ equals the sample space $\mathcal X$ of the physical parameter $X$, \eqref{AIN} is large for $A = \ell_X$, with $\P_0(A) = F_0(A)$ as in \eqref{FT} and $\P(A) = \delta_{x_0}(A)$ a one-point distribution at $x_0$, the observed value of $X$ (or the midpoint of $\ell_X$) \citep{DiazHossjer2022, ThorvaldsenHossjer2020}.

Based also on active information, a mathematical formalization of the epistemological notions of learning and knowledge acquisition was recently developed by \citet{HossjerDiazRao2022}, where knowledge is usually defined as ``justified true belief'' \citep{Gettier1963, IchikawaSteup2018, Schwitzgebel2021}. This means that a subject or agent $S$ \textit{knows} a proposition $p$ if 
\begin{itemize}
    \item[{\it i}] $S$ believes $p$, 
    \item[{\it ii}] $p$ is true, 
    \item[{\it iii}] $S$'s belief about $p$ is justified.
\end{itemize}
If only properties \textit{i} and \textit{ii} are satisfied, we say that $S$ \textit{learns} $p$. In other words, there is learning if and only if there is a true belief. Thus, there is a hierarchization from belief to knowledge through learning: 
\begin{align*}%\label{hierarchy}
    \text{knowledge} \subset \text{learning} \subset \text{belief}. 
\end{align*}
Moreover, the inclusions are proper, since it is possible to have a belief in a false proposition so that such a belief does not constitute learning; and it is also possible to learn $p$ without getting to know $p$, if the true belief cannot be justified. 

As for the mathematical formalization, belief is defined as a probability, as it is customary in Bayesian theory \citep[see, e.g.,][]{Berger2010, MacKay2003}, whereas propositions are parameters that have a true value. More explicitly, suppose we want to learn whether a proposition $p$ is either true or false. A set of possible worlds $\Omega$ is defined, where one world $\omega_0$ represents the value of the parameter $\omega$, referred to as true world, whereas all other $\omega\in\Omega\setminus\{\omega_0\}$ are counterfactuals. The set $A$ of interest is made of the worlds in which proposition $p$ is true. An ignorant person assigns beliefs to every subset of $\Omega$ according to some initial distribution  $\P_0$, whereas an agent $S$ with some data $D$ and \textbf{discernment} $\mathcal G$ (corresponding to a $\sigma$-algebra generated by the non-trivial events which data cannot discern into smaller events) updates his beliefs to $\P$, according to Bayes's rule,
\begin{align}\label{Post}
	\P(A) \defeq \frac{L(D \mid A)\P_0(A)}{L(D)},
\end{align}
with $L(D|A)$ the likelihood of observing $D$ given $A$. Then learning of $p$ is defined as follows.

\begin{definition}[{\bf Learning}]\label{D:L}
I) Agent $S$ has learnt about proposition $p$, compared to an ignorant person, either when $p$ is true and the posterior belief $\P$ about $p$ is higher than the prior belief $\P_0$ about $p$, or when $p$ is false and the posterior belief about $p$ is smaller than the prior belief about $p$. II) The agent $S$ has fully learnt $p$ (regardless of the beliefs of the ignorant person) if the posterior belief $\P$ about $p_1$ is 1 (0) when $p_1$ is true (false).  
\end{definition}

\begin{remark}\label{R:L}
Mathematically, the two parts of Definition \ref{D:L} can be phrased as follows: I) We say that there is learning about $p$, compared to an ignorant person, if
\begin{align}\label{Learning}
	\left\{
	\begin{array}{ll}
		0 < I^+(A), & \mbox{ and $p$ is true in the true world $\omega_0$},\\
		0> I^+(A), & \mbox{ and $p$ is false in the true world $\omega_0$}.
	\end{array}
	\right.
\end{align}
II) There is full learning about $p$ (regardless of the beliefs of the ignorant person) if $\P(A)=1$ ($\P(A)=0$) when $p$ is true (false) in the true world $\omega_0$.  
\end{remark}

In particular, a subject in maximum state of ignorance is represented by a maxent distribution $\P_0$ over $\Omega$.

Nonetheless, learning does not necessarily entail a particular belief about the true world, so it does not satisfy the conditions of a {\it justified} true belief, which requires having a belief for the right reasons. Knowledge acquisition is defined to cover this gap.

\begin{definition}[{\bf Knowledge}]\label{D:K}
I) Agent $S$ has acquired knowledge about $p$, compared to an ignorant person, if the following three conditions are satisfied: 
\begin{enumerate}
    \item $S$ has learnt about $p$, compared to the ignorant person.
    \item The true world $\omega_0$ is among the pool of possibilities for $S$, according to his posterior beliefs.
    \item The belief in $\omega_0$ under $\P$ is stronger than that under $\P_0$.
\end{enumerate}
II) Agent $S$ has acquired full knowledge about $p$ (regardless of the beliefs of the ignorant person) if $\P=\delta_{\omega_0}$.
\end{definition}

\begin{remark}
Mathematically, the three requirements of knowledge acquisition for Part I of Definition \ref{D:K} amount respectively to
\begin{enumerate}
    \item The criteria of Part I of Definition \ref{D:L} are satisfied.
    \item The true world $\omega_0$ is in $\text{supp}(\P)$, the support of $\P$.
    \item For all $\epsilon > 0$, the closed ball $B_\epsilon [\omega_0] \defeq \{\omega \in \Omega: d(\omega,\omega_0) \le \epsilon\}$ is such that $I^+(B_\epsilon[\omega_0])\ge0$, with strict inequality for some $\epsilon>0$, where $d$ is a metric over $\Omega$. 
\end{enumerate}
\end{remark}

Condition 1 ensures that knowledge is a more stringent concept than learning. Condition 2 is mathematically equivalent to saying that $S$ has a positive belief for every open ball centered at $\omega_0$ (i.e., if for all $\epsilon > 0$, $\P(B_\epsilon(\omega_0))>0$, where $B_\epsilon(\omega_0) \defeq \{\omega \in \Omega: d(\omega,\omega_0) < \epsilon\}$ is the open ball of radius $\epsilon$ centered at $\omega_0$), which in turn explains Condition 3.

\section{Learning and Knowledge in Cosmology}\label{S:LKFT}

It is pointless to think of different universes with different laws of nature because they are beyond our comprehension. Therefore, following the reasonable little question of \cite{Barnes2020}, we reduce the set of possible universes to those with the same laws but possibly different constants. In such scenario, there is a one-to-one relation between $\mathcal X$ and the multiverse because each $x\in\mathcal X$ represents a universe with value $x$ for the parameter $X$. In other words, there is a bijective function from $\mathcal X$ to the set of possible universes \citep{Adams2019, Sandora2019a, Sandora2019b, Sandora2019c, Sandora2019d, SandoraEtAl2022, SandoraEtAl2023a, SandoraEtAl2023b, SandoraEtAl2023c}, with $x_0$ being the actual value of the constant of nature {\it in our world}, and any other $x\in\mathcal X$ being a counterfactual universe. 

The framework from Section \ref{S:LKA} will be applied to analyze conditions under which FT can be known. As a preparation, we first consider the simpler proposition that our universe harbors life (which we know to be true).  

%\subsection{Learning and knowledge acquisition of a life permitting universe}\label{Sec:LPU} 

\begin{theorem}\label{T:p1}
    The proposition
    $$
        p_1: \mbox{ Our universe harbors life}
    $$
    can be fully learned and known. A sufficient condition for learning and knowledge acquisition of $p_1$, compared to an ignorant person, is having $\mathrm{TP}_{\max}<1$.
\end{theorem}

\begin{proof}
Identify  the set of possible worlds $\Omega$ for the learning and knowledge acquisition problem with the set of possible universes $\mathcal X$, with $\omega_0$ corresponding the observed value $x_0$ of $X$ for our universe. According to Section \ref{S:LKA}, the set $A=A_1$ of worlds for which $p_1$ is true coincides with $\ell_X$, the LPI of $X$. 

Let $\P$ be given by our best current theories and data. Since our theories and data admit knowledge of $\omega_0=x_0$ (i.e., we know that our universe admits life as well as the observed value $x_0$ of $X$), it follows that $\P=\delta_{x_0}$ has a one point distribution at $x_0$. Since $x_0\in\ell_X$, this implies 
\begin{equation}
\P(A_1) = \P(\ell_X) = 1. 
\label{PFTp1}
\end{equation}
Then, according to Part II) of Definition \ref{D:L} and Remark \ref{R:L}, we fully learn $p_1$ because $\P(A)=1$ and $p_1$ is true. Since $\P=\delta_{x_0}$, it follows from Part II) of Definition \ref{D:K} that we also acquire full knowledge about $p_1$. This completes the proof of the first part of Theorem \ref{T:p1}.

For a proof of the second part of Theorem \ref{T:p1}, we also need to consider the beliefs $\P_0$ of the ignorant person about the value of $X$. Recall that $\P_0$ is in maxent over $\Omega=\mathcal X$, given the restrictions imposed by $\mathcal F$, the class of distributions considered in Table \ref{T:Tuning}. This implies $\P_0\in\mathcal F$, and consequently
\begin{equation*}
\P_0(A_1) = \P_0(\ell_X) \le \max_{F\in {\mathcal F}} F(\ell_X) = \mathrm{TP}_{\max}.
\label{P0FTp1}
\end{equation*}
Assume $\mathrm{TP}_{\max}<1$. From this assumption and \eqref{P0FTp1} it follows that $\P_0(A_1)<\P(A_1)$, which, according to Part I) of Definition \ref{D:L} and Remark \ref{R:L} implies learning about $p_1$ compared to an ignorant person, since $p_1$ is true. The assumption $\mathrm{TP}_{\max}<1$ also implies knowledge acquisition about $p_1$ compared to an ignorant person, since Conditions 2 and 3 of Definition \ref{D:K} are satisfied as well.  
\end{proof}

\begin{remark}
    Notice that $p_1$ does not include the concept of fine-tuning. 
\end{remark}

After having dealt with learning and knowledge acquisition of our universe harboring life, let us now study whether it is possible to learn and know that our universe is fine-tuned. This corresponds to the proposition
$$
p_2: \mbox{ Our universe is fine-tuned}.
$$

We prove the following result:

\begin{theorem}\label{T:p2}
    Let $F_0$ be the actual distribution of $X$ and $0<\delta \ll 1$ the upper bound probability for fine-tuning, as stated in Definition \ref{D:FT}. Assume the agent $S$ believes that the assumption	
    \begin{align}\tag{A}\label{A}
	F_0\in\mathcal F
    \end{align}
    of Algorithm \ref{JesusLovesU} is true. The active information for the set of worlds $A$ for which $p_2$ is true satisfies 
    \begin{equation}
        I^+(A) = \min(\log\mathrm{TP}_{\mathrm{max}}^{-1},\delta^{-1}),
    \label{I+FT}
    \end{equation}
    from which 
    \begin{align*}
	I^+(A) \le  -\log \mathrm{TP}_{\mathrm{max}}
    \end{align*}
    follows. 
\end{theorem}
\begin{proof}
    Let $\Omega = (0,1]$ be the set of possible values of $F_0(\ell_X)$, whereas $\omega_0\in\Omega$ is the actual value of $F_0(\ell_X)$. Moreover, notice that $0\le F_0(\ell_X) \le \mathrm{TP}_\mathrm{\max}$ because of \eqref{A}. 
    %Together with \eqref{FT}, this implies that there is $\delta'\le\delta$ such that $F_0(\ell_X) \le \delta' \le \mathrm{TP}_\mathrm{\max}$. 
    It follows from Definition \ref{D:FT} that $A=A_2=(0,\delta]$ is the set of possible values of $F_0(\ell_X)$ that corresponds to a fine-tuned universe, that is, the set of values of $F_0(\ell_X)$ for which $p_2$ is true. 
    
     Since $\Omega$ is a bounded set, the maxent prior distribution $\P_0\sim U(0,1)$ is uniform. Consequently, $\P_0(A_2)=\delta$. Interpreting the agent's belief in \eqref{A} as new data $D$, and since $\P_0$ is uniform, it follows from \eqref{Post} that the posterior distribution $\P\sim U(0, \text{TP}_\text{max})$ is uniform as well. Therefore, $\P(A_2)=\min(\delta/\mathrm{TP}_\mathrm{\max},1)$. From this, it follows that  
\begin{equation*}
    I^+(A_2) = \log \frac{\min(\delta/\mathrm{TP}_\mathrm{\max},1)}{\delta} 
    = \log \min(\text{TP}_\text{max}^{-1},\delta^{-1}).  
\label{I+FT2}
\end{equation*}

\end{proof}

\begin{corollary}\label{LKFT}
    Assume that $p_2$ is true, so that \eqref{FT} is satisfied for some $0<\delta\ll 1$. Then    
    \begin{align}\label{TP}
	\mathrm{TP}_{\max} < 1.
    \end{align} 
    is enough to learn $p_2$, that $X$ is fine-tuned for life, whereas knowing $p_2$ requires more. For instance, \eqref{A} is a sufficient additional condition to assure that knowledge about $p_2$ has been acquired.
\end{corollary}

\begin{proof}
    Since $p_2$ is true by assumption, learning follows from Part I) of Definition \ref{D:L} and the fact that $I^+(A)>0$, which follows from \eqref{I+FT},  \eqref{TP} and $\delta < 1$. Note that although the agent's belief in \eqref{A} was used to construct the posterior belief $\P$ in Theorem \ref{T:p2}, it is not required that \eqref{A} is true to derive $\P$ and the formula in \eqref{I+FT} for $I^+(A)$.    
    
    If only \eqref{FT} and \eqref{TP}, but not \eqref{A}, hold, then it may happen that $\text{TP}_\text{max} <\omega_0 \le \delta$, so that for small enough $\epsilon>0$ we have $0=\P(B_\epsilon[\omega_0])<2\epsilon =\P_0(B_\epsilon[\omega_0])$. Whenever this happens, Condition 3 of Definition \ref{D:K} is violated. On the other hand, if \eqref{FT}, \eqref{TP}, and \eqref{A} hold, it is clear that $\omega_0\in (0,\text{TP}_\text{max})$. Since $\P_0\sim U(0,1)$ and $\P\sim U(0,\text{TP}_\text{max})$, it follows from Definition \ref{D:K} that the agent has acquired knowledge about $p_2$ compared to the ignorant person.   
\end{proof}

\begin{remark}
  Note that even though we assumed the agent's belief in \eqref{A} to construct $\P$, this condition is not needed for him to learn about FT. Comparing Theorem \ref{T:p1} with Corollary \ref{LKFT}, corresponding with intuition, we find that more is required to learn that our universe is fine-tuned ($p_2$) than to learn that it harbors life ($p_1$). 
\end{remark}

\begin{theorem}\label{SuffCond}
    \eqref{A} and
    \begin{align}\tag{B}\label{B}
        \text{TP}_{\max}\le\delta \ll 1
    \end{align}
    are sufficient for establishing, learning, and knowing that $X$ is fine-tuned for life.
\end{theorem}

\begin{proof}
    Since \eqref{A} and \eqref{B} imply \eqref{FT}, they are sufficient to establish FT (that $p_2$ is true). Also, \eqref{B} implies \eqref{TP}. Therefore, according to Corollary \ref{LKFT}, it follows that the agent has learnt and known FT compared to an ignorant person. 
\end{proof}
  
\begin{remark}\label{KFT}
    As seen in the proof of Corollary \ref{LKFT}, it is  possible to learn FT of $X$ for the wrong reasons if \eqref{FT} and \eqref{B} hold but \eqref{A} does not. However, this is unsatisfactory because it does not say anything about the real probability $\omega_0=F_0(\ell_X)$ of FT. Indeed, it is possible to falsely learn FT if \eqref{FT} is false while \eqref{TP} is true. Therefore, if FT cannot be known, learning is not useful. For this reason, we only pay attention to scenarios where FT can be known. 
\end{remark}

\section{On Knowing Fine-tuning}\label{S:Diff}

\subsection{The crucial choice of \texorpdfstring{$\mathcal F$}{F}}\label{S:CCF}

Theorem \ref{SuffCond} says that it is enough to focus on \eqref{A} and \eqref{B}, since these two equations imply that we can know that $X$ is fine-tuned for life. To analyse whether \eqref{A} and \eqref{B} are restrictive, recall that the first two steps of Algorithm \ref{JesusLovesU} determine the size of the set $\cal F$ of prior distributions of $X$, which represent possible beliefs of an ignorant person about our universe. It turns out that these two steps are of crucial importance for the validity of \eqref{A} and \eqref{B}, and hence for establishing that our universe is fine-tuned. The less we assume in Steps 1 and 2 of Algorithm \ref{JesusLovesU} (the larger $\mathcal F$ is), the easier it is to satisfy \eqref{A}, but the more difficult it is to satisfy \eqref{B}. This has interesting implications for the list of possible choices of $\mathcal F$ in Table \ref{T:Tuning}: 

First, since the maximal tuning probability $\text{TP}_{\max}$ is large for Rows 2, 5, and 8 of Table \ref{T:Tuning}, \eqref{B} is not satisfied and FT cannot be neither established nor known in these cases.

On the other hand, Rows 3, 4, 6, and 7 of Table \ref{T:Tuning} represent a second scenario where $\mathcal F$ is smaller, making it easy to satisfy \eqref{B} but more difficult to satisfy \eqref{A}. For instance, when the signal-to-noise ratio of the distribution $F_X$ of $X$ is bounded (Rows 3, 6, and 7), it is still possible that \eqref{A} does not hold (e.g., in the $\mathcal X=\mathbb R$ case, if the true maxent distribution $F_0$ of $X$ is positive, with a very small standard deviation in relation to its mean). 

\subsection{Random distributions}\label{S:RD}
For the second scenario of Section \ref{S:Diff}.\ref{S:CCF}, when $\mathcal F$ is chosen so small that \eqref{B} holds but \eqref{A} does not, one may assume that the distribution $F=F_X$ of $X$ is random. Under appropriate conditions, this leads to a generalization of Theorem \ref{SuffCond}, whereby FT of $X$ can be known, not with certainty but still with a high probability. 

In more detail, since $F_X=F(\cdot;\theta)$, such an approach leads to a having a random hyperparameter $\theta$, with the true distribution $F_0=F(\cdot;\theta_0)$ of $X$ corresponding to an instantiated value $\theta_0$ of $\theta$. It follows form the proofs of Corollary \ref{LKFT} and Theorem \ref{SuffCond} that if \eqref{B} holds, we can know that $X$ is fine-tuned, with a high probability $1-P(E)$, if the event 
\begin{equation}
E=\{F_0(\ell_X)>\text{TP}_{\text{max}}\} \subset \{F_0 \notin \mathcal{F}\}=\{\theta_0\notin \Theta\}
\label{E}
\end{equation}
has low probability. (Note that $P(E)$ being small is less stringent than \eqref{A} failing with a small probability). Showing that $P(E)$ is small thus requires choosing the distribution of $F=F(\cdot;\theta)$, for which a distribution $G_\theta$ of $\theta$ is needed. This requires implementing Algorithm \ref{JesusLovesU} on a new class of hyperhyperparameters $\lambda$ that parametrize the distribution $G_\theta =G(\cdot;\lambda)$ of $\theta$, which in turn leads to new restrictions $\lambda\in\Lambda$ on the hyperhyperparameter. We thus end up in an endless loop \citep{Hofstadter1999}  because it requires showing that the hyperparameters $\theta$ themselves are tuned to obtain FT of the physical parameters $X$. In other words, we replace the FT question of the physical parameters $X$ with a tuning question of the hyperparameters $\theta$. 

\subsection{Bounded parameter space}\label{S:BP}

The strategy of Section \ref{S:Diff}.\ref{S:RD}, to regard $F_X$ as random, would be great if the parameter space $\mathcal X = [a,b]$ of $X$ were finite and \eqref{B} were satisfied for the chosen class $\mathcal{F}$ of distributions of $X$. If there are no asymmetric restrictions on $F$, it is natural to regard $[a,b]$ as a torus and assume that $F$ is translation invariant in the sense that $F(\cdot)$ and $F(\cdot - x)$ have the same distribution for all $x$. It can then be proven, %\citep{DembskiMarks2010}, 
as a consequence of Markov's Inequality, that the probability of the set in \eqref{E} satisfies
$$
P(E)=P[F_0(\ell_X)>\text{TP}_{\text{max}}]
\le \frac{|\ell_X|}{[(b-a)\text{TP}_{\text{max}}]},
$$ 
which is small if 
\begin{equation}
|\ell_X|\ll \text{TP}_{\text{max}}.
\label{lXTP}
\end{equation}
This implies that, for a finite parameter space $\cal X$, we can know that $X$ is fine-tuned, with a large probability $1-P(E)$, if $\cal F$ is chosen small enough so that \eqref{B} holds, but also large enough so that \eqref{lXTP} is valid. 

Since a finite parameter space is a favorite scenario for physicists (remember the discussion after Theorem \ref{D:FT}), it is worth exploring whether there are actual FT instances in which the parameter space is actually finite. However, current arguments in favor of this hypothesis are unconvincing \citep{ColyvanGarfieldPriest2005, McGrewMcGrewVestrup2001}. Therefore, we need to deal with $\mathcal X$ being unbounded, and  for such cases FT cannot be known unless some restrictions, such as \eqref{A} and \eqref{B}, are imposed.   

\subsection{The real line as parameter space}

The argument of Section \ref{S:Diff}.\ref{S:BP}, based on random measures, is unfortunately not straightforward to generalize when the parameter space $\mathcal X$ of $X$ equals $\mathbb R$. Rows 4 and 5 of Table \ref{T:Tuning} correspond to a scale family of distributions that typically are symmetric around the origin. However, the scenario of Row 4 does not consider the observed value $x_0$ of the parameter in our universe as an estimator of a location parameter for $F_X$, making it plausible to suggest that \eqref{A} is violated. Indeed, it is a basic principle to obtain maxent distributions, subjecting them to the available observations \citep{CelisKeswaniVishnoi2020, SinghVishnoy2014, WainwrightJordan2008, Suh2023}. Thus, defining $\mathcal F$ without including into it a random variable whose location parameter is given by the observed value of $X$ in our universe might be too restrictive.
%Otherwise, this scenario FT cannot be known.

On the other hand, the location-scale family with an upper bound $T$ of the signal-to-noise ratio (Row 7 of Table \ref{T:Tuning}) does include distributions with $x_0$ as mean parameter. It is possible then to know that $X$ is fine-tuned if the true distribution $F_0$ satisfies $\mbox{SNR}\le T$ (so that \eqref{A} holds), and relative half length $\epsilon$ of the LPI is small enough so that \eqref{B} holds. The assumption $\mbox{SNR}\le T$ cannot be derived from the maxent principle though, and it rather needs some other justification (such as simplicity).  

\subsection{Positive line as parameter space}
When $\mathcal X=\mathbb R^+$, it is possible to use Row 1 of Table \ref{T:Tuning}. Notice that this is a scale family of distributions over $\mathbb R^+$ that includes all the exponential distributions. This is important because the exponential distribution $F(\cdot;\theta)$ is maxent over all distributions restricted to have mean $\theta$. Other maxent distributions over $\mathbb R^+$ are possible under different constraints; however, they are less parsimonious because they require more restrictions. For instance, the $\chi^2$ distribution of $X$ has maxent distribution over all distributions on the nonnegative reals given restrictions on $E(X)$ and $E(\log X)$ \cite[see, e.g.,][]{ParkBera2009}, thus pertaining to Row 2 of Table \ref{T:Tuning}. Hence, the family of exponential distributions $\{F(\cdot;\theta)\}_{\theta \in \mathbb R^+}$ is the less restrictive family of maxent distributions over $\mathbb R^+$, and we can naturally include into it the exponential distribution with mean value corresponding to the observed value $x_0$ of the constant in our universe. Therefore, in this scenario it is plausible that \eqref{A} is satisfied and, whenever \eqref{B} holds as well, we can establish and know FT.  To give a more explicit formulation of \eqref{B} for the family of exponential distributions of $X$ (cf.\ Row 1 of Table \ref{T:Tuning}), we notice that the maximum tuning probability equals
\begin{equation}
\text{TP}_{\text{max}} = 2C_1\epsilon = 2(\max_{x>0} xe^{-x})\epsilon = 2e^{-1}\epsilon
\label{TPmaxExp}
\end{equation}
when $\epsilon >0$ is small. 

By the reasoning above, we have thus established the following important Principle:

\begin{principle}\label{Princ}
    The fine-tuning of the parameter $X$ can be known when: 
    \begin{enumerate}
    \item The constant can only take nonnegative values ($\mathcal X = \mathbb R^+$).
    \item The size of the LPI is small relative to the observed value of the constant in our universe ($2e^{-1}\epsilon \le \delta \ll 1$).
\end{enumerate}
\end{principle}

\section{Examples}\label{S:Ex}

We consider several examples of physical parameters $X$ to determine whether FT can be known for them. We will only treat cases in which the parameter cannot take negative values, and make use of Principle \ref{Princ}. This Section does not intend to be exhaustive. It rather intends to show two things. First, that FT can be known in a few instances. Second, that we can fail to know the level of tuning for a parameter if the relative half length $\epsilon$ of its LPI is not small. More bounds for the LPIs can be found for instance in \citep{Adams2019} and references therein. In particular, Table 2 of \citep{Adams2019} shows that many LPIs span over several orders of magnitude. This implies that FT cannot be known in those cases because $\epsilon$ is large and therefore (by Table \ref{T:Tuning}) \eqref{B} is not satisfied.

\subsection{Critical density of the universe}

According to \cite{Davies1982}, the LPI of the critical density of the universe $\rho_\text{crit}$ is
\begin{align*}
	\ell_{\rho_\text{crit}} &= \rho_\text{crit,0}\left[1-10^{-60}, 1+ 10^{-60}\right],
\end{align*}
where $\rho_\text{crit,0}$ is the observed critical density. Since the density cannot be negative, $\mathcal X = \mathbb R^+$. Taking $\epsilon = 10^{-60}$, we notice that $\epsilon \ll 1$. According to row 1 of Table \ref{T:Tuning}, \cite{DiazHossjerMarks2023} and \eqref{TPmaxExp}, we have that $\text{TP}_\text{max} = 2e^{-1}\times 10^{-60}$ is a small number. Thus, by Principle \ref{Princ}, we know that there is FT.

\subsection{Gravitational force}

According to \cite{Davies1982}, when observing the ratio $X$ of the gravitational constant $G$ to the contribution from vacuum energy to the cosmological constant $\Lambda_\text{vac}$, the LPI of the gravitational constant is 
\begin{align*}
	\ell_X = x_0\left[1-10^{-100},1+10^{-100}\right], 
\end{align*}
where $x_0$ is the observed value of $X$. Since gravitation is attractive, it cannot be negative, and therefore $X$ is positive as well (so that $\mathcal X = \mathbb R^+$). Since $\epsilon = 10^{-100}$, it follows that $\epsilon \ll 1$. Thus, according to Row 1 of Table \ref{T:Tuning}, \cite{DiazHossjerMarks2021} and \eqref{TPmaxExp}, the maximal tuning probability is $\text{TP}_\text{max} = 2e^{-1}\times 10^{-100}$. By Principle \ref{Princ}, we know that there is FT.

%Notice that some physicists have considered the possibility that the gravitational force can also be repulsive [for instance, in \cite{Barnes2012} and \cite{LewisBarnes2016}]. In this scenario, additional assumptions beyond Principle \ref{Princ} are required to know that $X$ is fine-tuned.

\subsection{Higgs Vacuum Expectation Value}\label{vev}

The Higgs vacuum expectation value $v$ is given by $v_0=2\times 10^{-17}\text m_\text P$, where $\text m_\text P$ is the Planck mass. According to \cite{Barnes2012}, its LPI (in units of $m_\text P$) is given by
\begin{align}\label{ellv}
	\ell_v = \left[0.78 \times 10^{-17}, 3.3 \times 10^{-17}\right] = [a,b].
\end{align}
If $v$ falls below the lower bound $a$, the proton becomes heavier than the neutron, causing hydrogen to be unstable and enabling electron capture reaction \citep{DamourDonogue2008}. If $v$ exceeds the upper bound $b$ of $\ell_v$, there is no nuclear binding. The Higgs vacuum expectation value is positive as it describes the average value of the Higgs field in the vacuum (i.e., $\mathcal X = \mathbb R^+$). Writing \eqref{ellv} as in \eqref{D0LPI}, we obtain
\begin{align*}
    \ell_v = \frac{a+b}{2}[1-\epsilon,1+\epsilon] = 2.04\times 10^{-17}[1-0.617,1+0.617].
\end{align*}
Since $\epsilon=0.617$ is not small, we cannot use \eqref{TPmaxExp} to calculate the maximum tuning probability as $\text{TP}_{\max}=2e^{-1}\times 0.617 = 0.454$. Instead, we make use of the fact that for an exponential distribution $F(\cdot;\theta)$, with mean $\theta$, the tuning probability of $\ell_v$ equals
$$
F(\ell_v;\theta) = \exp(-a/\theta) - \exp(-b/\theta).
$$
This tuning probability is maximized for $\theta=(b-a)/\log(b/a)$, with a maximal value
\begin{equation}
\text{TP}_{\text{max}} = \left(\frac{b}{a}\right)^{-a/(b-a)} - \left(\frac{b}{a}\right)^{-b/(b-a)} = 0.488. 
\label{TPv1}
\end{equation}
Thus, despite the smallness of $\ell_v$, FT cannot be known for $v$ because $\text{TP}_{\text{max}}$ is large and hence \eqref{B} is not satisfied.

\subsection{Amplitude of primordial fluctuations}

According to \cite{Rees2000}, the LPI of the amplitude $Q$ of the primordial fluctuations is
\begin{align*}
\ell_Q &= \left[10^{-6}, 10^{-5}\right]\\
&= Q_0 [1-0.818,1+0.818]\\
&= [a,b],
\end{align*}
with $Q_0=5.5\cdot 10^{-6}$ the mid point of $\ell_Q$.
Since the amplitude cannot be negative, $\mathcal X = \mathbb R^+$. Since $\epsilon=0.818$ is not small, we follow the procedure of the previous example and calculate the maximal tuning probability as
\begin{align*}
\text{TP}_{\max} = \left(\frac{b}{a}\right)^{-a/(b-a)} - \left(\frac{b}{a}\right)^{-b/(b-a)} = 0.697,
\end{align*}
in agreement with \cite{DiazHossjerMarks2021}. Since this number is not small, the FT of $Q$ cannot be known.

\subsection{Baryon-Photon Ratio}\label{BPR}

According to \cite{Adams2019}, the baryon-photon ratio $\eta $ has a value $\eta\sim 6\times 10^{-10}$, and it can sustain life in between the interval
\begin{equation}
	\ell_\eta = [6\times 10^{-13}, 6 \times 10^{-7}] = [a,b].
\end{equation}

Again, since $\epsilon$ is not small, we follow the procedure of the previous two examples to calculate the maximal tuning probability
\begin{equation}
    \text{TP}_{\max}
    = \left(\frac{b}{a}\right)^{-a/(b-a)} - \left(\frac{b}{a}\right)^{-b/(b-a)}
    = 0.9999852.
\end{equation}
Thus, since $\text{TP}_{\max}$ is not small, the level of tuning of $\eta$ cannot be known. 

\begin{remark}
It follows from Examples \ref{vev}-\ref{BPR} that the maximal tuning probability of the life-permitting interval $\ell_X=[a,b]$ of $X\in \mathcal X=\mathbb R^+$, with a class of exponential priors for $F_X$, is a function 
\begin{equation}
\text{TP}_{\text{max}} = h(r) = r^{-1/(r-1)}-r^{-r/(r-1)}
\label{TPmaxExp2}
\end{equation}
of the ratio $r=b/a=(1+\epsilon)/(1-\epsilon)$ between the right and left end points of $\ell_X$. It can be seen that $h:(1,\infty)\to (0,1)$ is a strictly increasing function of $r$, with $\lim_{r\to 1} h(r) =: h(1)=0$, $h^\prime(1)=e^{-1}$ and $\lim_{r\to\infty} h(r)=1$. From this, it follows that \eqref{TPmaxExp} is a special case of \eqref{TPmaxExp2}, which holds in the limit of small half relative sizes $\epsilon>0$ of $\ell_X$. Moreover, from Theorem \ref{SuffCond}, a sufficient condition to know about FT of $X$ on $\mathcal X=\mathbb R^+$ is 
\begin{equation}
\frac{b}{a} \le h^{-1}(\delta).
\label{SuffCond2}
\end{equation}
When $\delta >0$ is small, \eqref{SuffCond2} is essentially equivalent to the condition on $\epsilon>0$ in the second part of Principle \ref{Princ}. 
\end{remark}

\section{Conclusion}

Using the recent mathematization  of knowledge acquisition in \citep{HossjerDiazRao2022}, we formalize Principle \ref{Princ}, which establishes that the FT of a given physical parameter $X$ can be known at least if the following two conditions are satisfied:
\begin{enumerate}
    \item The constant $X$ can only take nonnegative values.
    \item The size of its life-permitting interval is small relative to the observed value $x_0$ of the constant in our universe.
\end{enumerate}

This latter proviso is extremely important since, as we saw in the examples, it is possible to have very small intervals whose size relative to $x_0$ is large. As a result, \eqref{B} is not satisfied and the level of tuning cannot be known. In other words, to know that a given constant is fine-tuned, it is the relative half size of the interval that must be small, not the interval itself. That is, $|\ell_X| \ll x_0$ is required in order to satisfy \eqref{B}. Our conclusion is sobering in the sense that many things that, in spite of the ado, are touted as fine-tuned cannot be known to be so. 

On the other hand, when Principle \ref{Princ} is satisfied, we can be sure that FT can be known. However, we also highlight that not satisfying \eqref{B} does not imply that the tuning is coarse; it only implies that FT cannot be known, even if \eqref{A} is satisfied, since it is still possible that the real probability $F_0(\ell_X)$ of tuning is small. Thus, when the maximal tuning probability $\text{TP}_{\max}$ is not small, the conclusion of coarse tuning is epistemological, not ontological.

\bibliography{daangapaBibliography.bib}{}

\begin{thebibliography}{}
\expandafter\ifx\csname natexlab\endcsname\relax\def\natexlab#1{#1}\fi
\providecommand{\url}[1]{\href{#1}{#1}}
\providecommand{\dodoi}[1]{doi:~\href{http://doi.org/#1}{\nolinkurl{#1}}}
\providecommand{\doeprint}[1]{\href{http://ascl.net/#1}{\nolinkurl{http://ascl.net/#1}}}
\providecommand{\doarXiv}[1]{\href{https://arxiv.org/abs/#1}{\nolinkurl{https://arxiv.org/abs/#1}}}

\bibitem[{Adams(2019)}]{Adams2019}
Adams, F.~C. 2019, Physics Reports, 807, 1,
  \dodoi{10.1016/j.physrep.2019.02.001}

\bibitem[{Barnes(2012)}]{Barnes2012}
Barnes, L.~A. 2012, Publications of the Astronomical Society of Australia, 29,
  529, \dodoi{10.1071/AS12015}

\bibitem[{Barnes(2019-2020)}]{Barnes2020}
---. 2019-2020, Ergo, 6, 1220, \dodoi{10.3998/ergo.12405314.0006.042}

\bibitem[{Barnes(2021)}]{Barnes2021}
---. 2021, in The Routledge Companion to Philosophy of Physics, ed. E.~Knox \&
  A.~Wilson (New York: Routledge)

\bibitem[{Barrow \& Tipler(1988)}]{BarrowTipler1988}
Barrow, J.~D., \& Tipler, F.~J. 1988, The Anthropic Cosmological Principle
  (Oxford: Oxford University Press)

\bibitem[{Berger(2010)}]{Berger2010}
Berger, J. 2010, Statistical Decision Theory and Bayesian Analysis, 2nd edn.
  (New York: Springer)

\bibitem[{Bernoulli(1713)}]{Bernoulli1713}
Bernoulli, J. 1713, Ars {C}onjectandi (Basel: Thurneysen Brothers)

\bibitem[{Bostrom(2002)}]{Bostrom2002}
Bostrom, N. 2002, Anthropic Bias: Observation Selection Effects in Science and
  Philosophy (New York: Routledge)

\bibitem[{Carter(1974)}]{Carter1974}
Carter, B. 1974, in Confrontation of Cosmological Theories with Observational
  Data, ed. M.~S. Longhair (Dordrecht: D. Reidel), 291--298,
  \dodoi{10.1017/S0074180900235638}

\bibitem[{Celis {et~al.}(2020)Celis, Keswani, \&
  Vishnoi}]{CelisKeswaniVishnoi2020}
Celis, L.~E., Keswani, V., \& Vishnoi, N.~K. 2020, International Conference on
  Machine Learning, 1349

\bibitem[{Colyvan {et~al.}(2005)Colyvan, Garfield, \&
  Priest}]{ColyvanGarfieldPriest2005}
Colyvan, M., Garfield, J.~L., \& Priest, G. 2005, Synthese, 145, 325,
  \dodoi{10.1007/s11229-005-6195-0}

\bibitem[{Damour \& Donoghue(2008)}]{DamourDonogue2008}
Damour, T., \& Donoghue, J.~F. 2008, Physical Review D, 78, 014014,
  \dodoi{10.1103/PhysRevD.78.014014}

\bibitem[{Davies(1982)}]{Davies1982}
Davies, P. 1982, The Accidental Universe (Cambridge: Cambridge University
  Press)

\bibitem[{Davies(2008)}]{Davies2008}
---. 2008, The Goldilocks Enigma: Why Is the Universe Just Right for Life? (New
  York: Mariner Books)

\bibitem[{Dembski \& {Marks II}(2009)}]{DembskiMarks2009b}
Dembski, W.~A., \& {Marks II}, R.~J. 2009, IEEE Transactions Systems, Man, and
  Cybernetics - Part A: Systems and Humans, 5, 1051,
  \dodoi{10.1109/TSMCA.2009.2025027}

\bibitem[{D\'iaz-Pach{\'o}n \& H{\"o}ssjer(2022)}]{DiazHossjer2022}
D\'iaz-Pach{\'o}n, D.~A., \& H{\"o}ssjer, O. 2022, Entropy, 24, 1323,
  \dodoi{10.3390/e24101323}

\bibitem[{D\'iaz-Pach{\'o}n {et~al.}(2021)D\'iaz-Pach{\'o}n, H{\"o}ssjer, \&
  {Marks II}}]{DiazHossjerMarks2021}
D\'iaz-Pach{\'o}n, D.~A., H{\"o}ssjer, O., \& {Marks II}, R.~J. 2021, Journal
  of Cosmology and Astroparticle Physics, 2021, 020,
  \dodoi{10.1088/1475-7516/2021/07/020}

\bibitem[{D{\'\i}az-Pach{\'o}n {et~al.}(2023)D{\'\i}az-Pach{\'o}n, H{\"o}ssjer,
  \& {Marks II}}]{DiazHossjerMarks2023}
D{\'\i}az-Pach{\'o}n, D.~A., H{\"o}ssjer, O., \& {Marks II}, R.~J. 2023,
  Foundations of Physics, 53, \dodoi{10.1007/s10701-022-00650-1}

\bibitem[{D\'iaz-Pach{\'o}n \& {Marks II}(2020)}]{DiazMarks2020a}
D\'iaz-Pach{\'o}n, D.~A., \& {Marks II}, R.~J. 2020, BIO-Complexity, 2020, 1,
  \dodoi{10.5048/BIO-C.2020.3}

\bibitem[{D{\'\i}az-Pach{\'o}n \& {Marks II}(2020)}]{DiazMarks2020b}
D{\'\i}az-Pach{\'o}n, D.~A., \& {Marks II}, R.~J. 2020, BIO-Complexity, 2020,
  1, \dodoi{10.5048/BIO-C.2020.4}

\bibitem[{D\'iaz-Pach{\'o}n {et~al.}(2020)D\'iaz-Pach{\'o}n, S{\'a}enz, \&
  Rao}]{DiazSaenzRao2020}
D\'iaz-Pach{\'o}n, D.~A., S{\'a}enz, J.~P., \& Rao, J.~S. 2020, Statistics \&
  Probability Letters, 161, 108742, \dodoi{10.1016/j.spl.2020.108742}

\bibitem[{D{\'\i}az-Pach{\'o}n {et~al.}(2019)D{\'\i}az-Pach{\'o}n, S{\'a}enz,
  Rao, \& Dazard}]{DiazEtAl2019}
D{\'\i}az-Pach{\'o}n, D.~A., S{\'a}enz, J.~P., Rao, J.~S., \& Dazard, J.-E.
  2019, Applied Stochastic Models in Business and Industry, 35, 376,
  \dodoi{10.1002/asmb.2430}

\bibitem[{Gettier(1963)}]{Gettier1963}
Gettier, E.~L. 1963, Analysis, 23, 121, \dodoi{10.2307/3326922}

\bibitem[{Helbig(2023)}]{Helbig2023}
Helbig, P. 2023, Foundations of Physics, 53, \dodoi{10.1007/s10701-023-00732-8}

\bibitem[{Hofstadter(1999)}]{Hofstadter1999}
Hofstadter, D.~R. 1999, G{\"o}del, Escher, Bach: an Ethernal Golden Braid (New
  York: Basic Books)

\bibitem[{Hom {et~al.}(2023)Hom, Yik, \& Monta{\~n}ez}]{HomYikMontanez2023}
Hom, C., Yik, W., \& Monta{\~n}ez, G.~D. 2023, in 2023 IEEE 10th International
  Conference on Data Science and Advanced Analytics (DSAA), 1--11,
  \dodoi{10.1109/DSAA60987.2023.10302643}

\bibitem[{Hossenfelder(2020)}]{Hossenfelder2020}
Hossenfelder, S. 2020, Lost in Math: How Beauty Leads Physics Astray (New York:
  Basic Books)

\bibitem[{Hossenfelder(2021)}]{Hossenfelder2019}
---. 2021, Synthese, 198, 3727, \dodoi{10.1007/s11229-019-02377-5}

\bibitem[{H\"ossjer {et~al.}(2023)H\"ossjer, D{\'\i}az-Pach{\'o}n, Chen, \&
  Rao}]{HossjerEtAl2023}
H\"ossjer, O., D{\'\i}az-Pach{\'o}n, D.~A., Chen, Z., \& Rao, J.~S. 2023, IEEE
  Transactions on Information Theory, Accepted,
  \dodoi{10.1109/TIT.2023.3327399}

\bibitem[{H\"ossjer {et~al.}(2022)H\"ossjer, D{\'\i}az-Pach{\'o}n, \&
  Rao}]{HossjerDiazRao2022}
H\"ossjer, O., D{\'\i}az-Pach{\'o}n, D.~A., \& Rao, J.~S. 2022, Entropy, 24,
  1469, \dodoi{10.3390/e24101469}

\bibitem[{Ichikawa \& Steup(2018)}]{IchikawaSteup2018}
Ichikawa, J.~J., \& Steup, M. 2018, in The Stanford Encyclopedia of Philosophy,
  ed. E.~N. Zalta (Stanford: {Metaphysics Research Lab, Stanford University})

\bibitem[{Lewis \& Barnes(2016)}]{LewisBarnes2016}
Lewis, G.~F., \& Barnes, L.~A. 2016, A Fortunate Universe: Life In a Finely
  Tuned Cosmos (Cambridge: Cambridge University Press),
  \dodoi{10.1017/9781316661413}

\bibitem[{Liu {et~al.}(2023)Liu, D{\'\i}az-Pach{\'o}n, Rao, \&
  Dazard}]{LiuEtAl2023}
Liu, T., D{\'\i}az-Pach{\'o}n, D.~A., Rao, J.~S., \& Dazard, J.-E. 2023, IEEE
  Transactions on Pattern Analysis and Machine Intelligence, 45, 4637,
  \dodoi{10.1109/TPAMI.2022.3195462}

\bibitem[{MacKay(2003)}]{MacKay2003}
MacKay, D. J.~C. 2003, Information Theory, Inference, and Learning Algorithms
  (Cambridge: Cambridge University Press)

\bibitem[{McGrew \& McGrew(2005)}]{McGrewMcGrew2005}
McGrew, L., \& McGrew, T. 2005, Philosophia Christi, 7, 423,
  \dodoi{10.5840/pc20057235}

\bibitem[{McGrew(2018)}]{McGrew2018}
McGrew, T. 2018, Quaestiones Disputatae, 8, 147, \dodoi{10.5840/qd2018828}

\bibitem[{McGrew {et~al.}(2001)McGrew, McGrew, \&
  Vestrup}]{McGrewMcGrewVestrup2001}
McGrew, T., McGrew, L., \& Vestrup, E. 2001, Mind, New Series, 110, 1027,
  \dodoi{10.1093/mind/110.440.1027}

\bibitem[{Monta{\~n}ez(2017{\natexlab{a}})}]{Montanez2017a}
Monta{\~n}ez, G.~D. 2017{\natexlab{a}}, PhD thesis, Carnegie Mellon University,
  Pittsburgh, PA

\bibitem[{Monta{\~n}ez(2017{\natexlab{b}})}]{Montanez2017b}
---. 2017{\natexlab{b}}, 2017 IEEE International Conference on Systems, Man,
  and Cybernetics (SMC), 477, \dodoi{10.1109/SMC.2017.8122651}

\bibitem[{Monta{\~n}ez {et~al.}(2021)Monta{\~n}ez, Bashir, \&
  Lauw}]{MontanezEtAl2021}
Monta{\~n}ez, G.~D., Bashir, D., \& Lauw, J. 2021, in Agents and Artificial
  Intelligence, ed. A.~P. Rocha, L.~Steels, \& J.~ven~den Herik (Cham:
  Springer), 332--353, \dodoi{10.1007/978-3-030-71158-0_16}

\bibitem[{Monta{\~n}ez {et~al.}(2019)Monta{\~n}ez, Hayase, Lauw, Macias,
  Trikha, \& Vendemiatti}]{MontanezEtAl2019}
Monta{\~n}ez, G.~D., Hayase, J., Lauw, J., {et~al.} 2019, in 2nd Australasian
  Joint Conference on Artificial Intelligence (AI 2019), ed. J.~Liu \&
  J.~Bailey (Cham: Springer), 277--288, \dodoi{10.1007/978-3-030-35288-2_23}

\bibitem[{Park \& Bera(2009)}]{ParkBera2009}
Park, S.~Y., \& Bera, A.~K. 2009, Journal of Econometrics, 150, 219,
  \dodoi{10.1016/j.jeconom.2008.12.014}

\bibitem[{Rees(2000)}]{Rees2000}
Rees, M.~J. 2000, Just Six Numbers: The Deep Forces That Shape The Universe
  (New York: Basic Books)

\bibitem[{Sandora(2019{\natexlab{a}})}]{Sandora2019a}
Sandora, M. 2019{\natexlab{a}}, Universe, 5, 149,
  \dodoi{10.3390/universe5060149}

\bibitem[{Sandora(2019{\natexlab{b}})}]{Sandora2019b}
---. 2019{\natexlab{b}}, Universe, 5, 157, \dodoi{10.3390/universe5060157}

\bibitem[{Sandora(2019{\natexlab{c}})}]{Sandora2019c}
---. 2019{\natexlab{c}}, Universe, 5, 171, \dodoi{10.3390/universe5070171}

\bibitem[{Sandora(2019{\natexlab{d}})}]{Sandora2019d}
---. 2019{\natexlab{d}}, Universe, 5, 175, \dodoi{10.3390/universe5070175}

\bibitem[{Sandora {et~al.}(2023{\natexlab{a}})Sandora, Airapetian, Barnes, \&
  Lewis}]{SandoraEtAl2023a}
Sandora, M., Airapetian, V., Barnes, L.~A., \& Lewis, G.~F. 2023{\natexlab{a}},
  Universe, 9, 2, \dodoi{10.3390/universe9010002}

\bibitem[{Sandora {et~al.}(2023{\natexlab{b}})Sandora, Airapetian, Barnes, \&
  Lewis}]{SandoraEtAl2023b}
---. 2023{\natexlab{b}}, Universe, 9, 4, \dodoi{10.3390/universe9010004}

\bibitem[{Sandora {et~al.}(2022)Sandora, Airapetian, Barnes, Lewis, \&
  P{\'e}rez-Rodr{\'\i}guez}]{SandoraEtAl2022}
Sandora, M., Airapetian, V., Barnes, L.~A., Lewis, G.~F., \&
  P{\'e}rez-Rodr{\'\i}guez, I. 2022, Universe, 8, 651,
  \dodoi{10.3390/universe8120651}

\bibitem[{Sandora {et~al.}(2023{\natexlab{c}})Sandora, Airapetian, Barnes,
  Lewis, \& P{\'e}rez-Rodr{\'\i}guez}]{SandoraEtAl2023c}
---. 2023{\natexlab{c}}, Universe, 9, 42, \dodoi{10.3390/universe9010042}

\bibitem[{Schwitzgebel(2021)}]{Schwitzgebel2021}
Schwitzgebel, E. 2021, in The Stanford Encyclopedia of Philosophy, winter 2021
  edn., ed. E.~N. Zalta (Stanford: Metaphysics Research Lab, Stanford
  University)

\bibitem[{Singh \& Vishnoi(2014)}]{SinghVishnoy2014}
Singh, M., \& Vishnoi, N.~K. 2014, in Proceedings of the Forty-Sixth Annual ACM
  Symposium on Theory of Computing (New York: Association for Computing
  Machinery), 50--59, \dodoi{10.1145/2591796.2591803}

\bibitem[{Suh(2023)}]{Suh2023}
Suh, C. 2023, Information Theory for Data Science (Boston-Delft: Now
  Publishers)

\bibitem[{Tegmark(2015)}]{Tegmark2015}
Tegmark, M. 2015, Our Mathematical Universe (New York: Vintage)

\bibitem[{Tegmark {et~al.}(2006)Tegmark, Aguirre, Rees, \&
  Wilczek}]{TegmarkEtAl2006}
Tegmark, M., Aguirre, A., Rees, M., \& Wilczek, F. 2006, Physical Review D, 73,
  023505, \dodoi{10.1103/PhysRevD.73.023505}

\bibitem[{Thorvaldsen \& H{\"o}ssjer(2020)}]{ThorvaldsenHossjer2020}
Thorvaldsen, S., \& H{\"o}ssjer, O. 2020, Journal of Theoretical Biology, 501,
  110352, \dodoi{10.1016/j.jtbi.2020.110352}

\bibitem[{Thorvaldsen \& H{\"o}ssjer(2023)}]{ThorvaldsenHossjer2023}
---. 2023, Journal of the Royal Statistical Society Series C: Applied
  Statistics, qlad062, \dodoi{10.1093/jrsssc/qlad062}

\bibitem[{Wainwright \& Jordan(2008)}]{WainwrightJordan2008}
Wainwright, M.~J., \& Jordan, M.~I. 2008, Foundations and Trends in Machine
  Learning, 1, 1, \dodoi{10.1561/2200000001}

\bibitem[{Wolpert \& MacReady(1995)}]{WolpertMacReady1995}
Wolpert, D.~H., \& MacReady, W.~G. 1995, {No Free Lunch Theorems for Search},
  Tech. Rep. SFI-TR-95-02-010, Santa Fe Institute

\bibitem[{Wolpert \& MacReady(1997)}]{WolpertMacReady1997}
---. 1997, IEEE Transactions on Evolutionary Computation, 1, 67,
  \dodoi{10.1109/4235.585893}

\bibitem[{Zhou {et~al.}(2023)Zhou, D{\'\i}az-Pach{\'o}n, Zhao, Rao, \&
  H{\"o}ssjer}]{ZhouEtAl2023}
Zhou, L., D{\'\i}az-Pach{\'o}n, D.~A., Zhao, C., Rao, J.~S., \& H{\"o}ssjer, O.
  2023, Statistics in Medicine, 42, 4713, \dodoi{10.1002/sim.9885}

\end{thebibliography}
\bibliographystyle{aasjournal}

%\end{document}

%\bibliography{sample631}{}
%\bibliographystyle{aasjournal}



\end{document}